\documentclass[11pt,a4paper]{article}
\usepackage{jheppub}
\usepackage[utf8]{inputenc}
\usepackage[english]{babel}
\usepackage{amsfonts}
\usepackage{tcolorbox}[most] 

\usepackage[bottom]{footmisc}

\usepackage [autostyle]{csquotes}
\MakeOuterQuote{"}

\newcommand{\sgn}[1]{\mathrm{sgn}\!\left(#1\right)}

\newcommand{\sg}{\mathrm{sg}}
\newcommand{\sgtilde}{\tilde{\mathrm{sg}}}
\newcommand{\Li}[2]{\mathrm{Li}_{#1}\!\left(#2\right)}

\newcommand{\LiSV}[2]{\mathcal{L}_{#1}\!\left(#2\right)}
\newcommand{\eps}{\varepsilon}
\newcommand{\dx}{\mathrm{d}}
\newcommand{\e}{\mathrm{e}}
\newcommand{\iu}{\mathrm{i}}
\newcommand{\iz}{\mathrm{i}0}
\newcommand{\iztilde}{\mathrm{i}\tilde{0}}
\newcommand{\hyg}[1]{{_2}\mathrm{F}_1\!\left(#1\right)}

\title{The  massless non-adjacent double off-shell scalar box integral --- branch cut structure and all-order epsilon expansion}

\author[]{Juliane Haug}
\author[1]{and Fabian Wunder\note{Corresponding author}}

\affiliation[]{Institut f\"ur Theoretische Physik, Universit\"at T\"ubingen, \\
	Kepler Center for Astro and Particle Physics, \\ 
	Auf der Morgenstelle 14, D-72076 T\"ubingen, Germany}

\emailAdd{juliane-clara-celine.haug@uni-tuebingen.de}
\emailAdd{fabian.wunder@uni-tuebingen.de}

\abstract{We generalize the result of our recent paper on the massless single off-shell scalar box integral to the case of two non-adjacent end points off the light cone. An analytic result in $d=4-2\eps$ dimensions is established in terms of four Gauss hypergeometric ${_2}\mathrm{F}_1$ functions respectively their single-valued counterparts. This allows for an explicit splitting of real and imaginary parts, as well as an all-order $\eps$-expansion in terms of single-valued polylogarithms.}
\arxivnumber{2302.01956}

\begin{document}
\maketitle

%\flushbottom
                                   
%\clearpage 

\section{Introduction}	
The recent success of finding an all-order $\eps$-expansion for the single off-shell scalar box integral in terms of single-valued polylogarithms \citep{Haug2022} raised the question whether a similar result can be obtained in the case of more than one end point off the light cone.
Here, $\eps$ is the dimensional regularization parameter, with $d=4-2\eps$ the number of space-time dimensions.
In this paper, we show that a direct generalization of the method proposed in ref.\,\citep{Haug2022} to two particles off the light cone is indeed possible if they are on non-adjacent corners of the box. This results in a hypergeometric representation which allows for an explicit splitting of real and imaginary parts, as well as an all-order $\eps$-expansion. So far, the expansion was only available up to order $\eps^0$ in the literature \citep{BernDixonKosower1,DuplancicNizic,Ellis,Tarasov:2019mqy}.

For adjacent particles off the light cone a similar straightforward generalization does not seem feasible. Also in this case, results up to order $\eps^0$ have been known in the literature for some time \citep{BernDixonKosower1,DuplancicNizic,Ellis}. To calculate higher orders in the $\eps$-expansion for the adjacent case, as well as for more general kinematics, additional techniques will need to be employed.
The methods of differential equations \citep{Gehrmann:1999as} or Mellin-Barnes integrals \citep{Smirnov:1999gc}, which are highly successful in multi-loop calculations, could be considered.
A different approach towards the massless box integral via negative dimensions is outlined in ref.\,\citep{Anastasiou:1999cx}.
Yet another possibility are recurrence relations with respect to $d$, which were employed in ref.\,\citep{Fleischer:2003rm} to obtain a hypergeometric representation of the box integral for general kinematics and masses. Remarkably, a one-fold integral representation is presented there in eq.\,(96).
Using a wider set of functional equations, the box integral with massless propagators and arbitrary external masses in terms of Appell functions $\mathrm{F}_1$ and Gauss hypergeometric functions ${_2}\mathrm{F}_1$ was determined in ref.\,\citep{Tarasov:2019mqy} (see eqs.\,(4.29) and (4.61) of that reference). Comparing eqs.\,(4.29) and (4.32) of ref.\,\citep{Tarasov:2019mqy} furthermore suggests that the adjacent double off-shell box integral is of the same complexity as the general case.
The method of functional equations has recently been generalized to arbitrary internal masses \citep{Tarasov:2022clb}.
These results might be suitable starting points for generalizing the program of the present paper towards an all-order $\eps$-expansion with explicit real and imaginary parts for kinematics beyond the scope of the present work. 

Since the calculation for two non-adjacent particles off the light cone is very much analogous to the one presented in ref.\,\citep{Haug2022} for the single off-shell case, we kept the presentation more concise and refer the reader to our previous paper for further details.
This work is organized as follows.
In section \ref{sec:Calculating_D0}, we show that we can keep two non-adjacent particles off the light cone in the calculation of the box integral while closely following the steps of our previous work. We obtain a representation in terms of four Gauss hypergeometric functions ${_2}\mathrm{F}_1$, one more compared to the single off-shell case. This procedure introduces spurious branch cuts.
Since the occurring ${_2}\mathrm{F}_1$ functions are exactly of the kind present in the single-massive case, we can use their expansion known from ref.\,\citep{Haug2022} for the all-order $\eps$-expansion. In section \ref{sec:All order epsilon expansion}, we establish the all-order $\eps$-expansion of the non-adjacent double off-shell box integral, with real and imaginary parts made explicit. The result is free of spurious branch cuts due to the use of single-valued polylogarithms. 

\section{Calculating the non-adjacent double off-shell scalar box integral}\label{sec:Calculating_D0}
	\begin{figure}[b]
		\centering
			\includegraphics[width=0.4\textwidth]{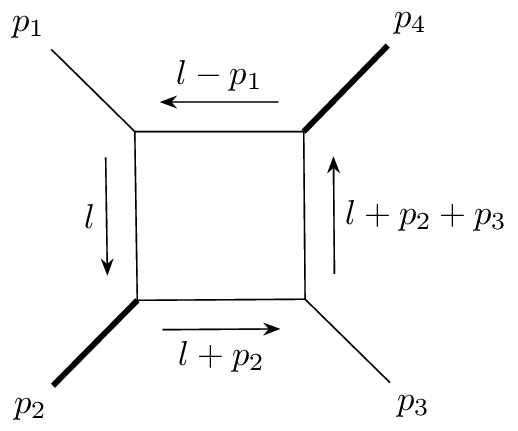}
		\caption{The scalar box diagram with two off-shell external particles diagonally opposite from each other carrying momenta $p_2$ and $p_4$. It is $p_1^2=p_3^2=0$ and $p_2^2=p_4^2\neq 0$. All external momenta are taken to be incoming. Drawn with Ti\textit{k}Z-Feynman \citep{Ellis2017}.}\label{fig:scalar_box_diagram}
	\end{figure}
\noindent{}The general scalar box integral with massless internal lines is given by
	\begin{align}
		\mathrm{D}_0 \,\equiv\, \frac{\mu^{4-d}}{\iu \pi^{d/2}} \int\dx^dl\,
		\frac{1}{\left[l^2 + \iu 
		 0\right] \left[(l+p_2)^2 +\iz\right] \left[(l+p_2+p_3)^2 +\iz\right] \left[(l-p_1)^2 +\iz\right]} \,,
	\end{align}
where the external momenta are labelled as indicated in the Feynman diagram depicted in figure \ref{fig:scalar_box_diagram}. We work in dimensional regularization with $d=4-2\eps$. While the starting point is the same as eq.\,(2.1) of our recent paper \citep{Haug2022}, here we consider the possibility of more than one external particle off the light cone. We have again explicitly kept the causal $+\iz$ in the propagators, which is necessary to determine the physical side of the branch cuts that appear in the final result.

After Feynman parametrization and evaluating the loop integral, we arrive at the integral representation (see also eq.\,(7) of ref.\,\citep{DuplancicNizic})
	\begin{align}
		D_0 \,&=\,\mu^{2\eps}\,\Gamma(2+\eps)\int_0^1 \frac{\dx x_1\, \dx x_2\, \dx x_3\, \dx x_4\,\delta(1-x_1-x_2-x_3-x_4)}{\left[-x_1 x_2 p_2^2 -x_1 x_3 s_2 -x_1 x_4 p_1^2 -x_2 x_3 p_3^2 -x_2 x_4 s_1 -x_3 x_4 p_4^2 - \iz\right]^{2+\eps}} \,,
		\label{eq:D0_FeynmanParametrized}
	\end{align}
with the Mandelstam variables $s_1=(p_1+p_2)^2$ and $s_2=(p_2+p_3)^2$. For convergence we require $\eps<0$. In the end, the result can be analytically continued to larger $\eps$.

Using the substitution from our previous paper (see also refs.\,\citep{Smirnov_Feynman_Integrals,FabiValeryAlexey}),
	\begin{align}
		x_1 \,&=\, \eta_1\xi_1\,, \quad x_2\,=\, \eta_2(1-\xi_2)\,, \quad
		x_3 \,=\, \eta_2\xi_2\,, \quad x_4 \,=\, \eta_1(1-\xi_1)\,,
		\label{eq:Substitution}
	\end{align}
the term in the denominator becomes
\begin{align}
&-\eta_1\eta_2(1-\xi_1)(1-\xi_2)\,s_1
\,-\,\eta_1\eta_2\xi_1\xi_2\,s_2
\,-\,\eta_1\eta_2 \xi_1(1-\xi_2)\,p_2^2
\,-\,\eta_1\eta_2\xi_2(1-\xi_1)\,p_4^2
\nonumber\\
&
-\,\eta_1^2\xi_1(1-\xi_1)\,p_1^2
\,-\,\eta_2^2\xi_2(1-\xi_2)\,p_3^2\, -\iz\,.
\label{eq:eta substitution}
\end{align}
Here, we see that the $\eta$-integrals factorize from the $\xi$-integrals if $p_1^2=p_3^2=0$. Hence, we can generalize the calculational method put forward in the single off-shell case to the case of two non-adjacent particles off the light cone.

We note that adjacent particles off the light cone require a different treatment. This is because setting for example $p_1^2=p_2^2=0$ in eq.\,\eqref{eq:D0_FeynmanParametrized} leads to a less symmetric Feynman parameter integral, where $x_1$ appears only once, $x_2$ and $x_4$ twice, and $x_3$ thrice. Therefore a substitution in the style of \eqref{eq:eta substitution} does not factorize the integral. Presumably, establishing an all order $\eps$-expansion in the case of adjacent particles off the light cone is considerably more difficult and requires additional techniques such as functional equations \citep{Tarasov:2019mqy}, differential equations \citep{Gehrmann:1999as} or Mellin-Barnes integrals \citep{Smirnov:1999gc}.
In the literature, it was also noted that the adjacent integral is "hard" in comparison to the "easy" non-adjacent integral \citep{BernDixonKosower1, DuplancicNizic}.
For this reason we only discuss the non-adjacent case in this work.

By evaluating the $\eta$-integrals in terms of Gamma functions and substituting $\xi_1\rightarrow 1-\xi_1$, we arrive at
\begin{align}
&D_0(s_1,s_2,p_1^2=0,p_2^2,p_3^2=0,p_4^2)\,=\,\mu^{2\eps}\frac{\Gamma(2+\eps)\Gamma^2(-\eps)}{\Gamma(-2\eps)}
\nonumber\\
&\times\int_0^1\dx\xi_1\int_0^1\dx\xi_2
\left[-p_2^2-(s_1-p_2^2)\xi_1-(s_2-p_2^2)\xi_2-(p_2^2+p_4^2-s_1-s_2)\xi_1\xi_2-\iz\right]^{-2-\eps},
\end{align}
which in the $p_2^2\rightarrow 0$ limit reduces to eq.\,(2.11) of our previous work \citep{Haug2022}. Note that $s_3\equiv(p_1+p_3)^2=p_2^2+p_4^2-s_1-s_2$. Analogous to the procedure there, we factor out $\left(\frac{(s_1-p_2^2)(s_2-p_2^2)}{p_2^2+p_4^2-s_1-s_2}-\iz\right)^{\!-2-\eps}$. This leads to
\begin{align}
D_0(s_1,s_2,0,p_2^2,0,p_4^2)=\frac{\Gamma(2+\eps)\Gamma^2(-\eps)}{\Gamma(-2\eps)}\left(\frac{\mu^2(p_2^2+p_4^2-s_1-s_2)}{(s_1-p_2^2)(s_2-p_2^2)}+\iz\right)^{\!\eps}
\frac{I_{123}(x_1,x_2,x_3)}{(s_1-p_2^2)(s_2-p_2^2)}\,,
\end{align}
where we introduced the dimensionless variables\footnote{To keep the notation close to our previous work \citep{Haug2022}, we use $x_{1,2,3}$ for the dimensionless variables. They are not to be confused with the Feynman parameters, which only appear in eqs.\,\eqref{eq:D0_FeynmanParametrized} and \eqref{eq:Substitution}.}
\begin{align}
x_1\equiv-\frac{p_2^2+p_4^2-s_1-s_2}{s_1-p_2^2}\,,\quad
x_2\equiv-\frac{p_2^2+p_4^2-s_1-s_2}{s_2-p_2^2}\,,\quad
x_3\equiv-\frac{(p_2^2+p_4^2-s_1-s_2)p_2^2}{(s_1-p_2^2)(s_2-p_2^2)}\,,
\end{align}
and
\begin{align}
I_{123}(x_1,x_2,x_3)=x_1 x_2 \int_0^1\dx\xi_1\int_0^1\dx\xi_2\left[x_3+x_2\xi_1+x_1\xi_3-x_1x_2\xi_1\xi_2+\iz\,\sg\right]^{-2-\eps}\,,
\end{align}
with $\sg\equiv\sgn{\frac{(s_1-p_2^2)(s_2-p_2^2)}{p_2^2+p_4^2-s_1-s_2}}$.
The integral $I_{123}(x_1,x_2,x_3)$ can be evaluated along the same lines as $I_{12}(x_1,x_2)$ in the single off-shell case. 
As a generalization of eq.\,(2.22) of ref.\,\citep{Haug2022} we find
\begin{align}
I_{123}(x_1,x_2,x_3)=\frac{(1+x_3-\iz\,\sg)^{-1-\eps}}{1+\eps}&\left[I\left(\frac{x_1+x_3}{1+x_3}\right)+I\left(\frac{x_2+x_3}{1+x_3}\right)
\right.
\nonumber\\
&\left.-I\left(\frac{x_1+x_2+x_3-x_1 x_2}{1+x_3}\right)-I\left(\frac{x_3}{1+x_3}\right)\right],
\label{eq:I_123 in terms of Ichi}
\end{align}
where
\begin{align}
I(\chi)\equiv\int_0^\chi\frac{\dx\zeta}{1-\zeta}\left(\left[\zeta-\iz\,\sgtilde\right]^{-1-\eps}-1\right),
\label{eq:Definition Ichi}
\end{align} 
with $\sgtilde=\sgn{1+x_3}\,\sg$.
We can evaluate $I(\chi)$ in terms of a hypergeometric function and an additional $\ln(1-\chi-\iztilde)$ as in eq.\,(2.26) of ref.\,\citep{Haug2022},
\begin{align}
		I(\chi) \,&=\, -\frac{1}{\eps} \left[\chi -\iz\, \sgtilde\right]^{-\eps} \hyg{1,-\eps,1-\eps;\chi+\iztilde} +\, \ln(1-\chi-\iztilde) \,. \label{eq:I_chi}
\end{align}
 This splitting introduces a new regulator $\iztilde$ (see ref.\,\citep{Haug2022} for details), independently for each $I(\chi)$. The real parts of the logarithms cancel between the first two and second two $I(\chi)$.
To achieve cancellation of the imaginary parts as well, we have to choose the $\iztilde_i$ of the four integrals such that
	\begin{align}
		0 \,\overset{!}{=}\, & \iu\pi \left[-\sgn{\tilde{0}_1}\Theta \left(\frac{x_1-1}{1+x_3}\right)
		-\sgn{\tilde{0}_2} \Theta\left(\frac{x_2-1}{1+x_3}\right) \right.
		\nonumber\\
		&\hphantom{\iu\pi} \left.\;\, +\sgn{\tilde{0}_3} \Theta\left(-\frac{(1-x_1)(1-x_2)}{1+x_3}\right)
		+\sgn{\tilde{0}_4}\Theta\left(-\frac{1}{1+x_3}\right) \right],
		\label{eq:i0 condition}
	\end{align}
where $\Theta(x)$ is the Heaviside step function.
This leads to the conditions tabulated in table \ref{tab: i0 conditions}.
\begin{table}
	\centering
	\begin{tabular}{ccccl}
		\hline
		$x_1$ & $x_2$ & $x_3$ & &Condition on regulators
		\\
		\hline
		$<1$ & $<1$ & $>-1$ & &$0\overset{!}{=}0$, i.e. no condition
		\\
		$<1$ & $<1$ & $<-1$ & &$0\overset{!}{=}-\tilde{0}_1-\tilde{0}_2+\tilde{0}_3+\tilde{0}_4$
		\\
		$<1$ & $<1$ & $>-1$ & &$0\overset{!}{=}-\tilde{0}_2+\tilde{0}_3$
		\\
		$<1$ & $>1$ & $<-1$ & &$0\overset{!}{=}-\tilde{0}_1+\tilde{0}_4$
		\\
		$>1$ & $>1$ & $>-1$ & &$0\overset{!}{=}-\tilde{0}_1+\tilde{0}_3$
		\\
		$>1$ & $<1$ & $<-1$ & &$0\overset{!}{=}-\tilde{0}_2+\tilde{0}_4$
		\\
		$>1$ & $<1$ & $>-1$ & &$0\overset{!}{=}-\tilde{0}_1-\tilde{0}_2$
		\\
		$>1$ & $>1$ & $<-1$ & &$0\overset{!}{=}\tilde{0}_3+\tilde{0}_4$
		\\
		\hline
		\end{tabular}
	\caption{Conditions for the regulators of the four $I(\chi)$ functions in eq.\,\eqref{eq:I_123 in terms of Ichi} imposed by eq.\,\eqref{eq:i0 condition} in the various kinematic regions.}
	\label{tab: i0 conditions}
	\end{table}
We can satisfy these by choosing
\begin{align}
&\iu\tilde{0}_1\equiv \iu\tilde{0}\,\sgn{\frac{x_1-x_2}{1+x_3}}=\iu\tilde{0}\,\sgn{\frac{(p_2^2+p_4^2-s_1-s_2)(s_1-s_2)}{s_1 s_2-p_2^2 p_4^2}},
\quad
\iu\tilde{0}_2\equiv -\iu \tilde{0}_1\,,
\nonumber\\
&\iu\tilde{0}_3\equiv \iu\tilde{0}\,\sgn{\frac{x_1x_2-x_1-x_2}{1+x_3}}=\iu\tilde{0}\,\sgn{\frac{(p_2^2+p_4^2-s_1-s_2)(p_4^2-p_2^2)}{s_1 s_2-p_2^2 p_4^2}},\quad
\iu\tilde{0}_4\equiv -\iu\tilde{0}_3\,.
\end{align}
For brevity, the common factor $\frac{p_2^2+p_4^2-s_1-s_2}{s_1 s_2-p_2^2 p_4^2}$ can be dropped in the signum expressions, since only relative signs matter.
Putting everyting together analogous to the calculation of the single off-shell integral in ref.\,\citep{Haug2022},
we find that the non-adjacent double off-shell scalar box integral is given by		
\begin{tcolorbox}[colback=green!10!white]
		\begin{align}
			D_0\!&\left(s_1,s_2,0,p_2^2,0,p_4^2\right) =\, \frac{1}{\eps^2} \frac{\Gamma(1+\eps)\, \Gamma^2(1-\eps)}{\Gamma(1-2\eps)}\, \frac{2}{s_1 s_2-p_2^2 p_4^2} \nonumber
			\\
			&\times \left\lbrace \left[\frac{\mu^2}{-s_1-\iz}\right]^{\eps} \hyg{1,-\eps,1-\eps;\,-\frac{(p_2^2+p_4^2-s_1-s_2)s_1}{s_1 s_2-p_2^2 p_4^2}+\iztilde\,\sgn{s_1-s_2}} \right. \nonumber
			\\
			&\left.\;\;\,+  \left[\frac{\mu^2}{-s_2-\iz}\right]^{\eps} \hyg{1,-\eps,1-\eps;\,-\frac{(p_2^2+p_4^2-s_1-s_2)s_2}{s_1 s_2-p_2^2 p_4^2}+\iztilde\,\sgn{s_2-s_1}}\right. \nonumber 
			\\
			&\left.\;\;\,- \left[\frac{\mu^2}{-p_2^2-\iz}\right]^{\eps} \hyg{1,-\eps,1-\eps;\,-\frac{(p_2^2+p_4^2-s_1-s_2)p_2^2}{s_1 s_2-p_2^2 p_4^2}+\iztilde\,\sgn{p_4^2-p_2^2}}\right. \nonumber 
			\\
			&\left.\;\;\,- \left[\frac{\mu^2}{-p_4^2-\iz}\right]^{\eps} \hyg{1,-\eps,1-\eps;\,-\frac{(p_2^2+p_4^2-s_1-s_2)p_4^2}{s_1 s_2-p_2^2 p_4^2}+\iztilde\,\sgn{p_2^2-p_4^2}} \right\rbrace . \label{eq:D0_general_result_combined_factors}
		\end{align}
\end{tcolorbox}
\noindent{}
This generalizes eq.\,(2.36) of ref.\,\citep{Haug2022}, where the single off-shell box integral is expressed in terms of three Gauss hypergeometric functions.
Remember that $s_3=p_2^2+p_4^2-s_1-s_2$.
Taking the limit $p_2^2\rightarrow 0$ or $p_4^2\rightarrow 0$ trivially reduces the result to the single off-shell box integral.\footnote{The branch cut of each hypergeometric function is the same in these limits if we restore the common factor $\frac{p_2^2+p_4^2-s_1-s_2}{s_1 s_2-p_2^2 p_4^2}$ in the signum expressions.} The result is explicitly symmetric under simultaneously interchanging $s_1\leftrightarrow s_2$ and $p_2^2\leftrightarrow p_4^2$, as well as under the interchange $s_1\leftrightarrow p_2^2$, $s_2\leftrightarrow p_4^2$.
The branch cuts of the individual hypergeometric functions are all spurious, i.e. they cancel in the sum.

While the hypergeometric representation of the single off-shell integral has been known in the literature for some time \cite{Matsuura1989,BernDixonKosower1}, a similar representation of the non-adjacent double off-shell box integral in terms of four Gauss hypergeometric functions was only recently found in ref.\,\citep{Tarasov:2019mqy} (see eqs.\,(4.34) and (4.66) of that reference). The main difference to our result is that by a suitable choice of the explicit $\iztilde_i$, we eliminated the logarithms which are present in eq.\,(4.66) of ref.\,\citep{Tarasov:2019mqy}.

\section{All order \boldmath{$\eps$}-expansion}
\label{sec:All order epsilon expansion}
To get rid of the spurious branch cuts introduced by the Gauss hypergeometric functions, we follow the strategy of section 3 of ref.\,\citep{Haug2022}. We use the single-valued counterparts of ${_2}\mathrm{F}_1$ introduced in eq.\,(3.15) of ref.\,\citep{Haug2022},
\begin{align}
\mathfrak{F}(\eps;x) \,\equiv\, 1 \,+\, \ln\!\left|\frac{x}{x-1}\right| \sum_{n=1}^\infty \frac{(-\eps)^n}{n!} \ln^{n-1}|x| \,-\, \sum_{n=2}^\infty\eps^n\LiSV{n}{x} \,,
			\label{eq:frakF definition}
\end{align}
with the single-valued polylogarithms (see Appendix C of ref.\,\citep{Haug2022}) defined by
\begin{align}
			\LiSV{n}{x}=\sum_{k=0}^{n-1}\frac{(-1)^k}{k!}\,\ln^k|x|\,\Li{n-k}{x}+\frac{(-1)^{n-1}}{n!}\ln^{n-1}|x|\ln|1-x|\,.
			\label{eq:Definiton SVP}
		\end{align}
The function $\mathfrak{F}$ is defined directly by its all-order $\eps$-expansion.
An alternative form useful for kinematic regularization purposes\footnote{For an example see Appendix D of ref.\,\citep{Haug2022}}, which resums the $\ln|x|$ terms, reads
\begin{align}
\mathfrak{F}(\eps;x)=|x|^{-\eps}+\eps\ln|1-x|-\sum_{n=2}^\infty \eps^n \left[\frac{(-1)^n\ln|1-x|\ln^{n-1}|x|}{n!}+\LiSV{n}{x}\right] .
\end{align}
Near $x=0$, $\mathfrak{F}(\eps;x)$ behaves like $|x|^{-\eps}$.
While the individual $\mathfrak{F}(\eps;x)$ have a $\ln|1-x|$ term, the divergence at $x=1$ cancels between the four $\mathfrak{F}(\eps;x)$ in the complete result.

Expressing the non-adjacent double off-shell scalar box integral in terms of $\mathfrak{F}$, we find the representation analogous to eq.\,(3.26) of ref.\,\citep{Haug2022},
\begin{tcolorbox}[colback=green!10!white]
\begin{align}
\!\!\!\!\!D_0\!\left(s_1,s_2,p_2^2,p_4^2\right)& =\, \frac{1}{\eps^2} \frac{\Gamma(1+\eps)\, \Gamma^2(1-\eps)}{\Gamma(1-2\eps)}\, \frac{2}{s_1 s_2-p_2^2 p_4^2} \left|\frac{(p_2^2+p_4^2-s_1-s_2)\mu^2}{s_1 s_2-p_2^2 p_4^2}\right|^\eps\nonumber
\\
			&\times \left\lbrace \left(\Theta(-s_1)+\Theta(s_1)\e^{\iu\pi\eps}\right) \,\mathfrak{F}\!\left(\eps;-\frac{(p_2^2+p_4^2-s_1-s_2)s_1}{s_1s_2-p_2^2p_4^2}\right) \right. \nonumber
			\\
			&\phantom{\times}+\left(\Theta(-s_2)+\Theta(s_2)\e^{\iu\pi\eps}\right) \,\mathfrak{F}\!\left(\eps;-\frac{(p_2^2+p_4^2-s_1-s_2)s_2}{s_1s_2-p_2^2p_4^2}\right)\nonumber
			\\
			&\phantom{\times}-\left(\Theta(-p_2^2)+\Theta(p_2^2)\e^{\iu\pi\eps}\right) \,\mathfrak{F}\!\left(\eps;-\frac{(p_2^2+p_4^2-s_1-s_2)p_2^2}{s_1s_2-p_2^2p_4^2}\right)\nonumber
			\\
			&\left.\!\phantom{\times}-\left(\Theta(-p_4^2)+\Theta(p_4^2)\e^{\iu\pi\eps}\right) \,\mathfrak{F}\!\left(\eps;-\frac{(p_2^2+p_4^2-s_1-s_2)p_4^2}{s_1s_2-p_2^2p_4^2}\right) \right\rbrace.
\label{eq:D0_general_result_Ffrak}
\end{align}
\end{tcolorbox}
\noindent{}Real and imaginary parts in every kinematic region, as well as the all-order $\eps$-expansion, can conveniently be read off in this form. Note that eq.\,\eqref{eq:D0_general_result_Ffrak} is free of spurious branch cuts due to the use of single-valued polylogarithms.  
This representation constitutes the main new result of this work.

\section{Conclusion}
We have generalized our recent result for the single off-shell scalar box integral to the case of two non-adjacent external particles off the light cone and established an all-order $\eps$-expansion with explicit real and imaginary parts. The solution turns out to be even more symmetric than the single off-shell case to which it trivially reduces in the limit of either one particle put on the light cone.
However, for most applications the adjacent double off-shell integral would also be needed. For that integral, a result in the style found for the non-adjacent case does not seem to be accessible by a straightforward generalization of the method proposed in ref.\,\citep{Haug2022}. This is related to the lower degree of symmetry in its Feynman parameter representation. The results of ref.\,\citep{Tarasov:2019mqy} found through functional equation techniques could be a promising starting point towards a similar all-order $\eps$-expansion of the adjacent double-off shell as well as the triple and fully off-shell cases.

\acknowledgments
We thank the anonymous referee of ref.\,\citep{Haug2022} for inspiring this paper by raising the question whether the result is generalizable to two end points off the light cone.
We also thank Oleg V. Tarasov for drawing our attention to the similar hypergeometric representation he obtained for the non-adjacent double off-shell box integral.
We are grateful to Werner Vogelsang for helpful comments.
This study was supported in part by Deutsche Forschungsgemeinschaft (DFG) through the Research Unit FOR 2926 (Project No. 409651613).
{J.~H.} is grateful to the Landesgraduiertenf{\"o}rderung Baden-W{\"u}rttemberg for supporting her research.

%%%%%%%%%%%%%%%%%%%%%%%%%%%%%%%%%%%%%%%%%%%%%%%%%%%%%%%%%%%%%%%%%%%%%%%%%%%%%%%%%%%%%%%
%
% The bibliography begins here
%
%%%%%%%%%%%%%%%%%%%%%%%%%%%%%%%%%%%%%%%%%%%%%%%%%%%%%%%%%%%%%%%%%%%%%%%%%%%%%%%%%%%%%%%
\bibliography{Bibliography_double_off_shell_scalar_box_integral}
\bibliographystyle{JHEP}
\end{document}